\documentclass[a4paper]{jpconf}
\usepackage{graphicx}
\begin{document}
\title{Fitting formulae for photon spectra from WIMP annihilation}

\author{J.\,A.\,R.\,Cembranos$^{1}$, A.\,de la Cruz-Dombriz$^{1,2}$,  A.\,Dobado$^{1}$, R.\,Lineros$^{3}$ and A.\,L.\,Maroto$^{1}$}
\address{
$^{1}$ Departamento de F\'{\i}sica Te\'orica I, Universidad Complutense de Madrid, E-28040, Spain.
 \\
$^{2}$ Department of Mathematics and Applied Mathematics, University of Cape Town, South Africa.
 \\
$^{3}$ 
IFIC, CSIC-Universidad de Valencia, Ed. Institutos, Apdo. correos 22085, E-46071, Spain.
}
\ead{dombriz at fis.ucm.es}

\begin{abstract}
Annihilation of different dark matter (DM) candidates into
Standard Model (SM) particles
could be detected through their contribution to the gamma ray fluxes
that are measured on the Earth.
The magnitude of such contributions
depends on the particular DM candidate, but certain
imprints of produced photon 
spectra
may be analyzed in a 
model-independent
fashion. 
In this work we provide the fitting formulae
for the photon spectra generated by WIMP 
annihilation into quarks, 
leptons and gauge bosons
channels in a wide range of WIMP masses. 
\end{abstract}

\section{Introduction}
According to present observations of large scale structures, CMB anisotropies 
and light nuclei abundances, 
DM cannot be accommodated within the SM of elementary particles.
Indeed, DM  is not only a required component on cosmological 
scales, but it is also needed to provide a 
satisfactory description of rotational speeds of galaxies,
orbital velocities of galaxies in clusters, gravitational lensing
of background objects by galaxy clusters
and the temperature distribution of hot gas in galaxies and
clusters of galaxies.
The experimental determination of the DM nature will require the
interplay of collider experiments 
and astrophysical
observations. Concerning the latter, DM searches use to be classified 
in direct or indirect ones (see [1] and references in the introduction
section in [2]). 
Concerning direct detection, the 
elastic scattering of DM particles from nuclei should lead directly 
to observable nuclear recoil signatures although
the weak interactions between DM and the 
standard matter makes DM direct detection extremely difficult.

On the other hand, DM might be detected indirectly, by observing
their annihilation products  into standard model particles. Thus, even if WIMPs
(Weakly Interacting Massive Particles) are stable, two of them may annihilate
into ordinary matter such as quarks
, leptons and gauge bosons. Their annihilation in different places (galactic halo, Sun, etc.)
produce cosmic rays which could be discriminated from the background
through distinctive signatures. A
cascade process follows the WIMPs annihilation.  
In the end, the potentially observable
stable particles would be neutrinos, gamma rays, positrons
and antimatter 
that may be observed through different devices. Among them, neutrinos 
and gamma rays have the advantage of maintaining their original
direction due to their null electric charges. 
%

Concerning gamma rays, photon fluxes in specific (mainly SUSY) DM models 
are usually obtained
by software packages such as DarkSUSY and micrOMEGAs 
based on PYTHIA Monte Carlo event generator [3] 
after having fixed a WIMP mass 
for the particular model under consideration. 

This communication precisely focuses 
on providing  
the fitting functions  for the photon production coming from
quarks (including $t$ quark), leptons 
and gauge bosons annihilations channels. This would allow to
apply the results to alternative DM candidates for which
software packages have not been developed.
In addition, our investigation determine the dependence of such spectra on the
WIMP mass in  a model independent way. 

On the other hand, the information about channel contribution and
mass dependence can be very useful in order to identify gamma-ray
signals for specific WIMP candidates and may also provide
relevant information about the photon energy distribution 
when SM particle pairs annihilate.

Let us remind that the $\gamma$-ray flux from the annihilation 
of two WIMPs of mass $M$ into two SM 
particles coming from all possible annihilation channels (labelled by the subindex $i$) is given by:
\begin{eqnarray}
\label{eq:integrand}
\frac{{\rm d}\,\Phi_{\gamma}^{{\rm DM}}}{{\rm d}E_{\gamma}}=\frac{1}{4\pi M^2}
 \sum_i\langle\sigma_i v\rangle
\frac{{\rm d}N_\gamma^i}{{\rm d}E_{\gamma}}
\;\, \times \;\,
 \frac{1}{\Delta\Omega} \int_{\Delta\Omega} {\rm d}\Omega
 \int_{\rm l.o.s.} \rho^2[r(s)]\ {\rm d}s\;,
\end{eqnarray}
%
%
where $\langle \sigma_{i} v \rangle$ holds for
the thermal averaged annihilation cross-section of two 
WIMPs into two ($i^{th}$ channel) SM 
particles and $\rho$ is the DM density.
The integral is performed along the line of sight (l.o.s.) to the
target and averaged over the detector solid angle $\Delta\Omega$.
%
%
%

%
%
%
%
\section{Procedure}

We have used the particle physics PYTHIA software 
[3] to obtain our results.
The WIMPs annihilation is usually 
split into two separated processes: The first describes the
annihilation of WIMPs and its SM output.
The second one considers the evolution 
of the obtained SM unstable products.
Due to the typical velocity dispersion of DM, we expect
most of the annihilations to happen quasi-statically. This fact 
allows to state that by considering different  
center of mass (CM) energies for the obtained SM particles pairs 
from WIMP annihilation process, we are
indeed studying different WIMP masses, i.e. $E_{{\rm CM}} \simeq 2\,M$. 
The procedure to obtain the photon spectra is thus straightforward: For a given pair of 
SM particles which are produced in the WIMP annihilation, we count the number of photons.
%
The number of simulated collisions in each bin was fixed in order to
provide suitable statistics
in the number of produced photons. For instance, for the high energy
bins many collisions are required to get a significant number of
photons,  whereas for low-intermediate energy, many photons are
usually produced even for a small number of collisions. 


\section{Results}

According to the extensive PYTHIA simulations
performed for electroweak gauge bosons, leptons and quarks channels,
three different fitting functions were required 
to fit the photon spectra ${ \rm d}N^{i}_{\gamma}/{\rm d}E_{\gamma}$: 
one for light quarks and leptons, one for gauge bosons ($W$ and $Z$) 
and a final one for $t$ quark. Those 
expressions depended on both WIMP mass dependent and independent
parameters. Concerning the WIMP mass independent parameters, their
values nevertheless do depend on the considered annihilation
channel whereas for WIMP mass dependent ones, their evolutions with 
WIMP mass are parametrized 
by continuous and smooth curves as seen in [2]. Figure 1 shows two of these fits. 
The resulting expressions for the fitting functions are the following:
%
\subsection{Light quarks and leptons}
For $q\bar{q}$ (except the $t\bar{t}$ studied separately
in section $3.3$), $\tau^{+}\tau^{-}$ and $\mu^{+}\mu^{-}$ channels,
the most general formula needed to reproduce the $
{\rm d}N^{i}_{\gamma}/{\rm d}E_{\gamma}$ simulations 
may be written as:

\begin{eqnarray}
\frac{{\rm d}N_{\gamma}}{{\rm d}x}\,=\,\frac{a_{1}}{x^{1.5}}\,
{\rm exp}
\left(-b_{1} x^{n_1}-b_2 x^{n_2} -\frac{c_{1}}{x^{d_1}} 
+\frac{c_2}{x^{d_2}}\right) + q\,
{\rm ln}
\left[p(1-x^{l})\right]\frac{x^2-2x+2}{x}
\label{general_formula}
\end{eqnarray}
where the logarithmic term
takes into account the final state radiation through a
Weizs\"{a}cker-Williams expression. 
%
Strictly speaking, the $p$ parameter in the
Weizs\"{a}cker-Williams term in the previous
formula is $(M/m_{particle})^2$
where $m_{particle}$ is  the mass of the
charged particle that emits radiation. However in our approach, it will be
a free parameter to be fitted since the radiation comes
from many possible charged particles, which are produced
along the decay and hadronization processes. Therefore
all the bremsstrahlung effects were encapsulated in a
single Weizs\"{a}cker-Williams-like term by using $p$ and $q$ parameters.  $l$ parameter is equal to one except for the $\mu^{+}\mu^{-}$
channel.

In fact, for the $\mu^{+}\mu^{-}$ channel, the above expression
(\ref{general_formula}) becomes simpler
since the exponential contribution is absent. 
The total gamma rays flux for this channel is thus well fitted by:

\begin{eqnarray}
\frac{{\rm d}N_{\gamma}}{{\rm d}x}\,=\, q\,{\rm ln}\left[p(1-x^{l})\right]\frac{x^2-2x+2}{x}
\label{general_formula_mu}
\end{eqnarray}
For both leptons channels the covered WIMP mass range 
was from $25$ to $5\cdot10^4$ ${\rm GeV}$. Concerning the 
mass dependent parameters,  they are $n_1$ and $p$ in expression 
(\ref{general_formula}) for $\tau^{+}\tau^{-}$ channel
(the rest of parameters are mass independent) 
and  $p$, $q$ and $l$ in expression (\ref{general_formula_mu}) for $\mu$ lepton.  For $q\bar{q}$
channels, no general rule about mass (in)dependence of parameters can be settled, but specific results 
channel by channel are presented in [2].

Let us finally mention that  the gamma rays from
$e^{-}e^{+}$ pairs, the only contribution is that coming from bremsstrahlung. Therefore,
the previous expression (\ref{general_formula_mu}) is also valid
with $q=\alpha_{{\rm QED}}/\pi$, $p=\left(M/m_{e^{-}}\right)^{2}$ and $l\equiv1$, parameters 
choice that obviously corresponds to the well-known
Weizs\"{a}cker-Williams formula.
\subsection{$W$ and $Z$ gauge bosons}
For the $W^{+}W^{-}$ and $ZZ$ channels, the parametrization used to
 fit the Monte Carlo simulation is:
\begin{eqnarray}
x^{1.5}\frac{{\rm d}N_{\gamma}}{{\rm d}x}\,=\, a_{1}\,{\rm exp}\left(-b_{1}\, x^{n_1}
-\frac{c_{1}}{x^{d_1}}\right)\left\{\frac{{\rm ln}[p(j-x)]}{{\rm ln}\,p}\right\}^{q}
\label{general_formula_W_Z}
\end{eqnarray}
This expression differs from expression (\ref{general_formula}) in
the absence of the additive logarithmic contribution that
acquires nonetheless a multiplicative behaviour. The exponential
contribution is also quite simple with only one positive and one
negative power laws. Moreover, $a_1$, $n_1$ and $q$ parameters
are WIMP mass independent (but they have different values depending on the studied 
gauge boson channel). The rest of parameters, i.e.,
$b_1$, $c_1$, $d_1$, $p$ and $j$ turned out to be both channel and mass dependent.
The covered WIMP mass range 
was from $100$ to $10^4$ ${\rm GeV}$ for both 
channels. Nonetheless, at masses higher
than 1000 ${\rm GeV}$, no significant change in the 
photon spectra for both channels was observed.

\subsection{$t$ quark}
Unlike the rest of the $q\bar{q}$ channels, the parametrization given by (\ref{general_formula})
is not valid when photon spectra are studied in the $t\bar{t}$ channel. For this one, the parametrization 
used to fit the simulations is:
\begin{eqnarray}
x^{1.5}\frac{{\rm d}N_{\gamma}}{{\rm d}x}\,=\, a_{1}\,{\rm exp}\left(-b_{1}\, x^{n_1}
-\frac{c_{1}}{x^{d_1}}-\frac{c_{2}}{x^{d_2}}\right)\left\{\frac{{\rm ln}[p(1-x^{l})]}{{\rm ln}\,p}\right\}^{q}
\label{general_formula_t}
\end{eqnarray}
%
%
In this formula, 
the exponential contribution is more complicated than the
one in expression (\ref{general_formula_W_Z}), with one positive and
two negative power laws. Again, the additive logarithmic contribution
is absent and acquires a multiplicative behaviour. Notice the 
exponent $l$ in the logarithmic argument, required to provide correct
fits in this channel.

The covered WIMP mass range 
was from $200$ to $10^5$ ${\rm GeV}$. Nevertheless, at masses higher than
1000 ${\rm GeV}$  no significant
change in the  gamma-ray spectra was observed. The mass dependent
parameters for this channel are $b_1$, $n_1$, $c_2$, $p$, $q$ and $l$ 
whereas $a_1$, $c_1$, $d_1$ and $d_2$ are mass independent. 
\section{Conclusions}

We have presented the model-independent fitting functions for the photon spectra coming from
WIMPs pair annihilation into Standard Model particle-antiparticle
pairs for all the phenomenologically relevant channels. 
Explicit calculations for all studied
channels [2] are available at the website
%
.
%
%
\begin{center}
\begin{figure}
\includegraphics[height=.225\textheight]{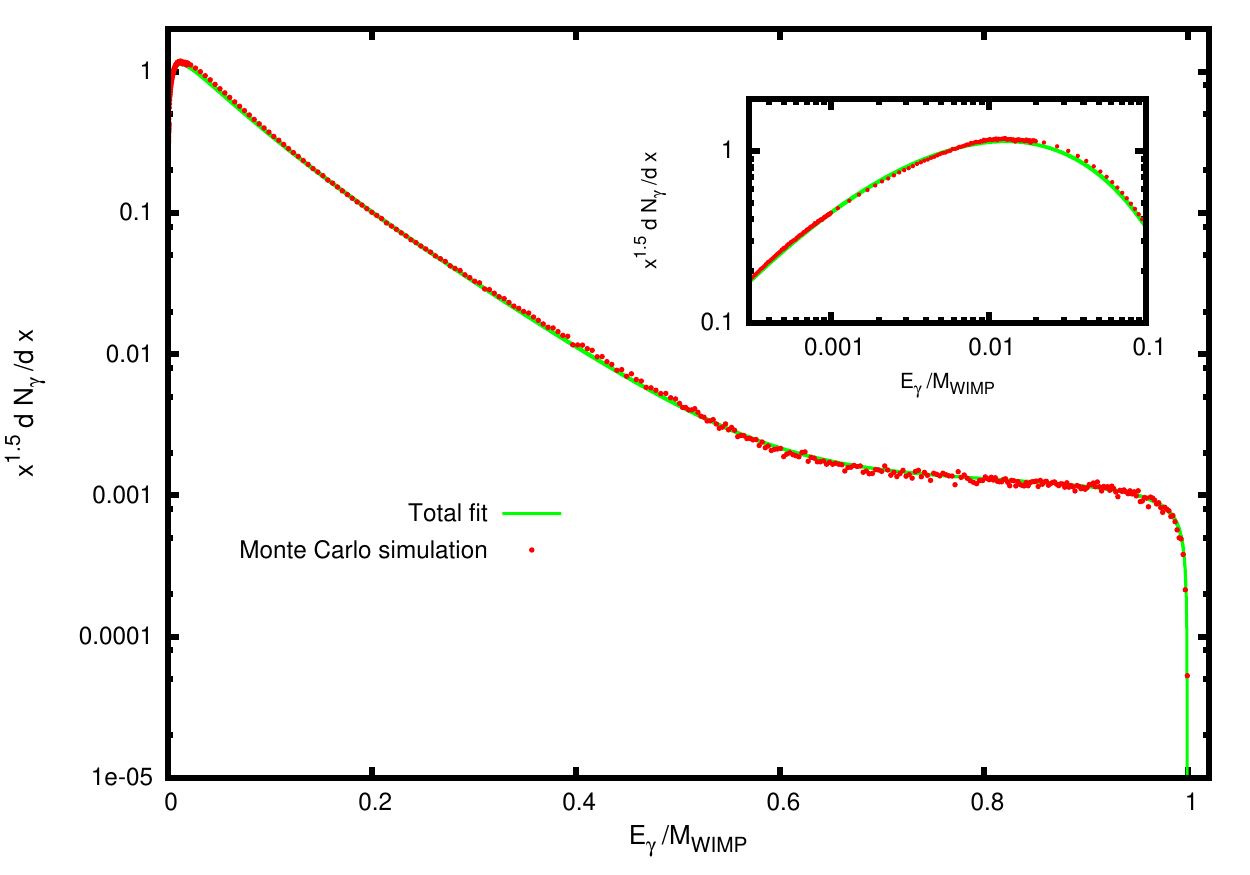}
\includegraphics[height=.225\textheight]{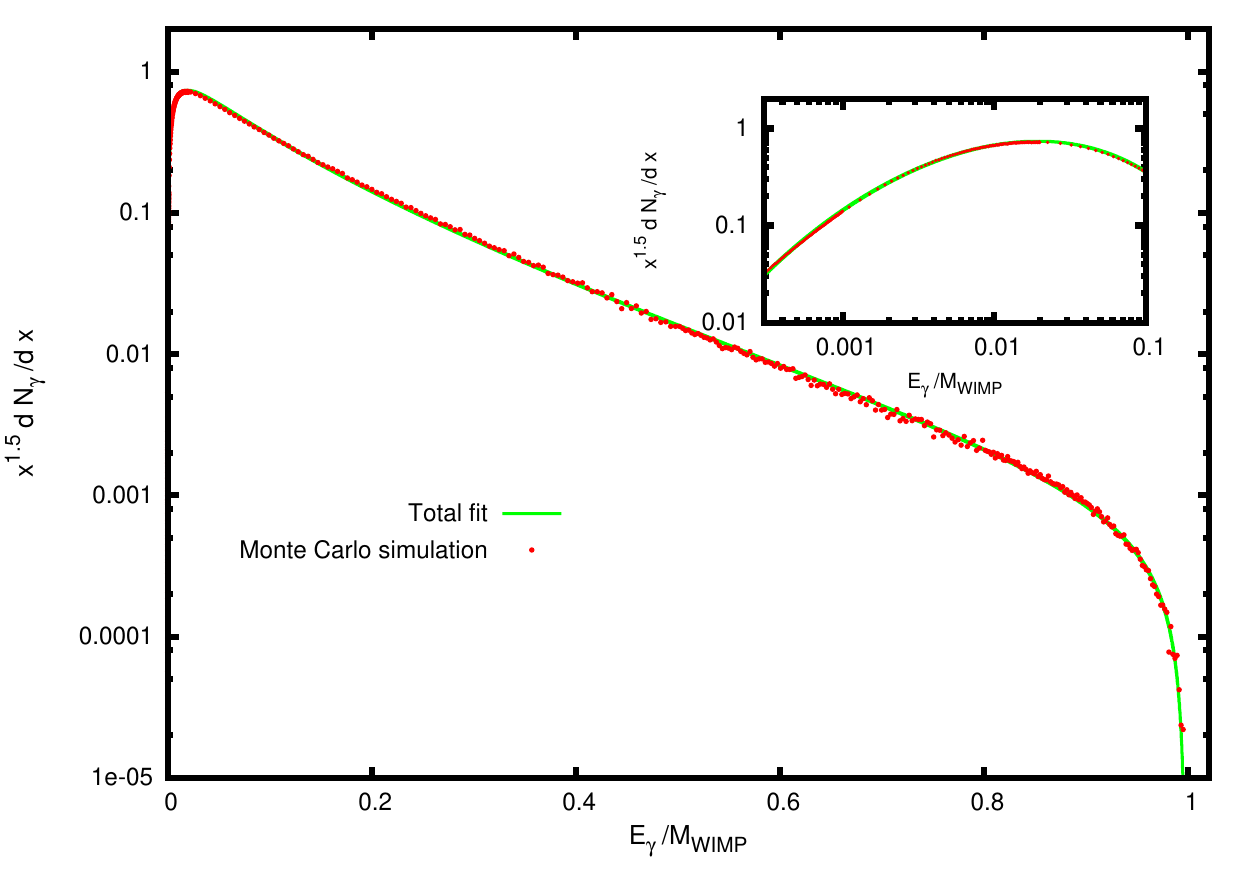}
\caption{Photon spectra for $M_{WIMP}=1$ TeV  in the $b\bar{b}$ (left) and $W^{+}W^{-}$ (right) channels.  
Dotted points are PYTHIA simulations and lines correspond to the proposed fitting functions.}
\end{figure}
\end{center}
%
%
%
%
%
\section{Acknowledgments}
This work was supported by MICINN (Spain) project numbers FIS 2008-01323 and 
FPA 2008-00592, FPA2008-00319, CAM/UCM 910309, 
and Consolider-Ingenio MULTIDARK CSD2009-00064. AdlCD was also supported 
by the National Research Foundation (South Africa). 
RL was also supported by the EC contract UNILHC PITN-GA-2009-237920
and PROMETEO/2009/091 (Generalitat Valenciana).

%


\smallskip
\end{document}